\documentclass[10pt]{article} 
\usepackage{graphicx,times,amsmath, amssymb}
\usepackage[T1]{fontenc} 
\usepackage{fixltx2e} 
\usepackage{xspace}

\begin{document}
\title{The Impact of the Topology on Cascading Failures in Electric Power Grids}

\author{Yakup Ko\c{c}$^{1,}$\thanks{Corresponding Author,~~Email: \texttt{Y.Koc@tudelft.nl},  Address: Jaffalaan 5, 2628BX Delft, The Netherlands, Phone: +31 (0)15 27 88380}~~Martijn Warnier$^1$~~Piet Van Mieghem$^{2}$\\~~Robert E. Kooij$^{2,3}$~~Frances M.T. Brazier$^1$\\
 $^1$Faculty of Technology, Policy and Management\\
 Delft University of Technology, Delft, the Netherlands\\
 $^2$ Faculty of Electrical Engineering, Mathematics and Computer Science\\
 Delft University of Technology, Delft, the Netherlands\\
 $^3$TNO (Netherlands Organisation for Applied Scientific Research)\\Delft, the Netherlands
 }
\date{}

\maketitle
\thispagestyle{empty}

\begin{abstract}
Cascading failures are one of the main reasons for blackouts in power transmission grids. The topology of a power grid, together with its operative state determine, for the most part, the robustness of the power grid against cascading failures. Secure electrical power supply requires, together with careful operation, a robust design of the electrical power grid topology. This paper investigates the impact of a power grid topology on its robustness against cascading failures. Currently, the impact of the topology on a grid robustness is mainly assessed by using purely topological approaches that fail to capture the essence of electric power flow. This paper proposes a metric, the effective graph resistance, that relates the topology of a power grid to its robustness against cascading failures by deliberate attacks, while also taking the fundamental characteristics of the electric power grid into account such as power flow allocation according to Kirchoff Laws. Experimental verification shows that the proposed metric anticipates the grid robustness accurately. The proposed metric is used to optimize a grid topology for a higher level of robustness. To demonstrate its applicability, the metric is applied on the IEEE 118 bus power system to improve its robustness against cascading failures.

\end{abstract}

\section{Introduction}
\label{sec_Introduction}
	Electric power is indispensable to modern societies. Security and availability of power supply is crucial. Disruption of power delivery systems has severe effects on the public order and substantial economic cost for society~\cite{Hines2009}. The safe operation of the power grid greatly reduces the risk of large-scale blackouts. Many countries, however, suffer from catastrophic blackouts paralysing daily life~\cite{BrazilBlackoutRef, blackoutReport}. Analysis of international blackout data reveals that the probability distribution of the blackout sizes decreases with the size of the blackout in an approximate power law regime~\cite{Dobson2007_2} with an exponent between $-1$ and $-2$, i.e.\xspace doubling the blackout size approximately halves its probability suggesting large scale blackouts are more likely than expected. 	
	
Cascading failures are the consequence of the collective dynamics of the complex power grid. Large scale cascades are typically due to propagation of a local failure into the global network. Consequently, analysing and mitigating cascading failures requires a system level approach. Recent advances in the field of network science~\cite{Barabasi99, Watts98,Buldyrev2010, Huang2013} reveal the promising potential of complex networks theory to investigate power grids vulnerability at a system level.
	
	This paper considers the \emph{vulnerability} of a system as a sensitivity to threats (i.e.\xspace malicious attacks) and disturbances (e.g.\xspace random failures) that possibly limit the ability of the system to accomplish its tasks, and provide the intended services. As the polar opposite of vulnerability, \emph{robustness} refers to the ability of a system to avoid malfunctioning when a fraction of its elements fail~\cite{Casals2007}.

	
	
	The operative state (e.g.\xspace loading level and power flow distribution), the structural aspects of a power grid (e.g.\xspace type of buses and their interconnection), and design choices of engineering systems (e.g.\xspace the protection systems, automatic controls, and towers and insulators) determine the robustness against cascading failures in a power grid. Increasing a power grid robustness requires a careful assessment of engineering system design choices and optimization of the operative state and the topology of the grid. The operative state of a power grid is continuously changing while the structure remains mainly unchanged. Optimization of the operative state of a power grid is a short-term optimization problem and requires a dynamic optimization approach~\cite{Koc2013, Koc2013_2, Pournaras2013}. On the other hand, optimization of the topology is a long-term optimization problem and requires investigating the impact of the grid topology on cascading failures. This paper focuses on the optimization of the topology, and investigates the relationship between a power grid topology and its robustness against cascading failures.
	

 One way to analyse the impact of the structural aspects of a power grid on the cascading effect is investigating the relationship between the robustness level against cascading failures and the general network properties of power grids. The electric power grid has small-world characteristics~\cite{Watts98, Kim2007}. A small average shortest path length (together with high clustering coefficient~\cite{Mieghem2011}) is one of the key characteristics of the small-world phenomenon. In small-world networks, the average shortest path length might dominate the network dynamics including, for example disease spreading~\cite{Kim2007}. These studies/results motivate power system researchers to investigate the impact of average shortest path length on the cascading failures robustness in power grids~\cite{Kim2007, Chen2006}. They deploy a purely topological approach and assume that electric power behaves as a discrete data packet and follows the shortest or the most efficient path between two nodes. Relying on this assumption, the average shortest path length is determined, and its impact on the network robustness is investigated. However, this purely topological approach does not comply with the fundamental characteristics of power grids. Electric power obeys the laws of Kirchoff and flows through all available paths rather than following a distinct path (e.g.\xspace shortest path) like e.g.\xspace a data packet in a data communication network. Therefore, the notion of distance needs to be tailored based on power systems fundamentals. This paper proposes the \emph{effective resistance} as a measure of electrical path length between two nodes, and the \emph{effective graph resistance} as a metric referring to electrical average path length of a power grid. The effective graph resistance~\cite{Mieghem2011} relates to the impact of the topology on cascading failure robustness while accounting for the fundamental properties of power grids such as power flow distribution through the grid according to Kirchoff Laws.

\section{Dynamic Model of Cascading Failures in Power Grids}
\label{sec_Model}

A power grid is a three-layered complex interconnected network consisting of generation, transmission, and distribution parts. Electric power is shipped from the generation buses to distribution substations through the transmission buses, all interconnected by transmission lines. Electric power flows in a grid according to Kirchoff's laws. Accordingly, impedances, voltage levels at each individual power station, voltage phase differences between power stations and loads at terminal stations control the power flow in the grid. AC power flow equations are non-linear equations that approximate the flows of both active and reactive powers. DC load flow equations are a linearised version of the AC power flow equations considering only flow of active power~\cite{Hertem2006}. DC power flow analysis introduces an average line flow error up to 5\% while it is 7 to 10 times faster than AC load flow model~\cite{Hertem2006}. Following Ref.~\cite{Dobson2007, Bao2009, Kinney2004}, this paper deploys DC load flow analysis to estimate the flow values across the network.

The maximum capacity (i.e.\xspace flow limit) of a line is defined as the maximum power flow that can be afforded by the line. The flow limit of a transmission line is imposed by thermal, stability or voltage drop constraints~\cite{Glover2001}. As many other recent studies~\cite{Motter2002, Motter2004, Crucitti2004, Bao2009, Ma2013, Wei2012, Chen2009, Chen2010}, this paper assumes that the maximum capacity of a line relates to its base (i.e. initial) load as follows:
\begin{equation}
\label{eq:capEstimation}
C_{i}=\alpha_{i} L_{i}(0)
\end{equation}

\noindent where \emph{$C_{i}$} is the maximum capacity, \emph{$L_{i}(0)$} is the base load, and $\alpha_{i}$~\cite{Motter2002} is the tolerance parameter of line \emph{$i$}. In a power grid, each line has a relay protecting it from permanent damage due to e.g.\xspace excessive flows. For instance, in case of overloading, an over-current relay notifies a circuit breaker to trip a line when the current of the line exceeds its rated limit (i.e.\xspace maximum capacity) and this violation lasts long enough to permanently damage the line. For the sake of simplicity, this paper assumes a deterministic model for the line tripping mechanism, i.e.\xspace a circuit breaker for a line trips at the moment the flow of the line exceeds its rated limit. While the over-current~\cite{Zhang2006} relays are not the only relays that are used in the power transmission grid, this paper models the other type of relaying mechanisms (e.g. distance and differential protection) as over-current relays.

An initial outage of a component changes the balance of the power flow distribution over the grid and causes a redistribution of the power flow over the network. This dynamic response of the system to this triggering event might overload other parts in the network. The protection mechanism trips these newly overloaded components, and the power flow is again redistributed potentially resulting in new overloads. In case of islanding, cascading failures continue in each island in which generators or loads are shed respectively to attain a power balance. The cascade of failures continues until no more components are overloaded. After the cascade subsides, the robustness of the grid against cascading failures is quantified in terms of the fraction of the served power demand ($DS$) after the cascading failures.

\section{Effective Graph Resistance in Electric Power Grids}
\label{sec_Graph Resistance Basics}
This section explains the relevant basic concepts from complex networks theory, explains the effective graph resistance, and elaborates on how it is computed in electric power grids. 

\subsection{\label{app_preliminaries}Complex networks preliminaries}
A network \emph{G(N,L)} consisting of a set of nodes \emph{N} and links \emph{L}, can fully be represented by its adjacency matrix $A$. The \emph{adjacency matrix} of a simple, unweighted graph \emph{G(N,L)} is an $N \times N$ symmetric matrix reflecting the interconnection of the nodes in the graph: $a_{ij}=0$ indicates that there is no edge, otherwise $a_{ij}=1$. In case of a weighted graph, the network is represented by a weighted adjacency matrix $W$ where $w_{ij}$ corresponds to the weight of the line between nodes $i$ and $j$; a weight can be a distance, cost, or delay. 

The \emph{Laplacian matrix}~\cite{Mieghem2011} $Q$ is another way to fully characterize a graph, and is defined as:
\begin{equation}\label{Laplacian}
Q=\Delta - A 
\end{equation}

\noindent where $\Delta$ is the diagonal matrix of the strengths of $G$: $\delta_{i}$=$\sum_{j}^{N}{w_{ij}}$. Hence, the Laplacian can be constructed as follows: 
\begin{equation}
  Q_{ij}=\begin{cases}
      
    \delta_{i}, 	& \text{if $i=j$}.\\
     -w_{ij},			& \text{if $i \neq j$ and $(i,j) \in L$}\\
     0,				& \text{otherwise}.

  \end{cases}
\end{equation}

\noindent where $\delta_{i}$ is the strength of node $\emph{i}$, and \emph{L} is the set of links in \emph{G}.

A \emph{path} $P_{ij}$ between pair of nodes \emph{i} and \emph{j} refers to the set of links connecting these nodes. The \emph{path length} $l(P_{ij})$ is the sum of the weights of constituent edges in the path $P_{ij}$. The shortest path length $l(P^{*}_{ij})$ is the minimizer of $l(P_{ij})$ over all $P_{ij}$. The \emph{average shortest path length} $l_{G}$ of a network $G$ is defined as:
\begin{equation}\label{CharPathLength}
l_{G}=\frac{1}{N(N-1)}\sum_{\substack{i \neq j \in G}}{l(P^{*}_{ij})}
\end{equation}
      
\noindent where \emph{N} is the number of nodes in the graph.


\subsection{\label{app_Computation of effective graph resistance}Effective graph resistance and its computation in electric power grids}
The effective resistance~\cite{Mieghem2011} $R_{ij}$ between a pair of nodes $i$ and $j$ is the potential difference between these nodes when a unit current is injected at node $i$ and withdrawn at node $j$. And the effective graph resistance $R_{G}$ is the sum of the individual effective resistance between each pair of nodes in the network. Computation of the effective graph resistance for a power grid necessitates information about the topology of the grid (i.e.\xspace interconnection of nodes), and reactance (or susceptance~\cite{Grainger1994}) values of the transmission lines in the grid. The effective graph resistance can be computed in two different ways: (a) by aggregating the effective resistances between each pair of nodes, and (b) by the eigenvalues of the Laplacian matrix of the grid.

The required steps to compute the effective graph resistance based on pairwise effective resistances are (i) constructing the Laplacian matrix of the grid, (ii) determining the generalised inverse of the Laplacian matrix, (iii) computing effective resistances between each pair of nodes, and (iv) summing up the effective resistances. 

The Laplacian matrix of a power grid $Q$ reflects the interconnection of buses with transmission lines. The weight $w_{ij}$ corresponds to the susceptance (i.e.\xspace inverse of reactance) value between nodes \emph{i} and \emph{j}. The Laplacian matrix constructed by the susceptance values is equivalent to the admittance matrix in the electrical power systems theory.


The \emph{effective resistance} $R_{ij}$ between any pair of nodes $i$ and $j$ is computed as:
\begin{equation}\label{EffResistance}
R_{ij}=Q_{ii}^{+}-2Q_{ij}^{+}+Q_{jj}^{+}
\end{equation}

\noindent where $Q^{+}$ is the Moore-Penrose pseudo-inverse of the $Q$.


The \emph{effective graph resistance} $R_{G}$ of a power network is then computed by summing up all the effective resistances between all pairs in a network. 
\begin{equation}\label{EffGraphResistance}
R_{G}={\sum_{i=1}^{N}}{\sum_{j=i+1}^{N} {R_{ij}}}
\end{equation}


Another way to compute the effective graph resistance of a power grid requires computation of the eigenvalues of the Laplacian matrix of the grid. This approach requires summing the inverse of the eigenvalues:
\begin{equation}\label{EffGraphResistanceEigenValues}
R_{G}=N{\sum_{i=1}^{N-1}} \frac{1}{\mu_{i}}
\end{equation}

 \noindent where $\mu_{i}$ is the $i^{th}$ eigenvalue of the Laplacian matrix, and $\mu_{1} \geq \mu_{2} \geq ...\geq \mu_{N-1} \geq \mu_{N}$. This methodology is computationally more efficient, but it does not give any insight into the individual electrical path lengths between pair of buses.

\section{\label{sec_Graph Resistance}Effective Graph Resistance as a Robustness Metric}
\begin{figure}[htb]
\centering
\includegraphics[scale=0.40]{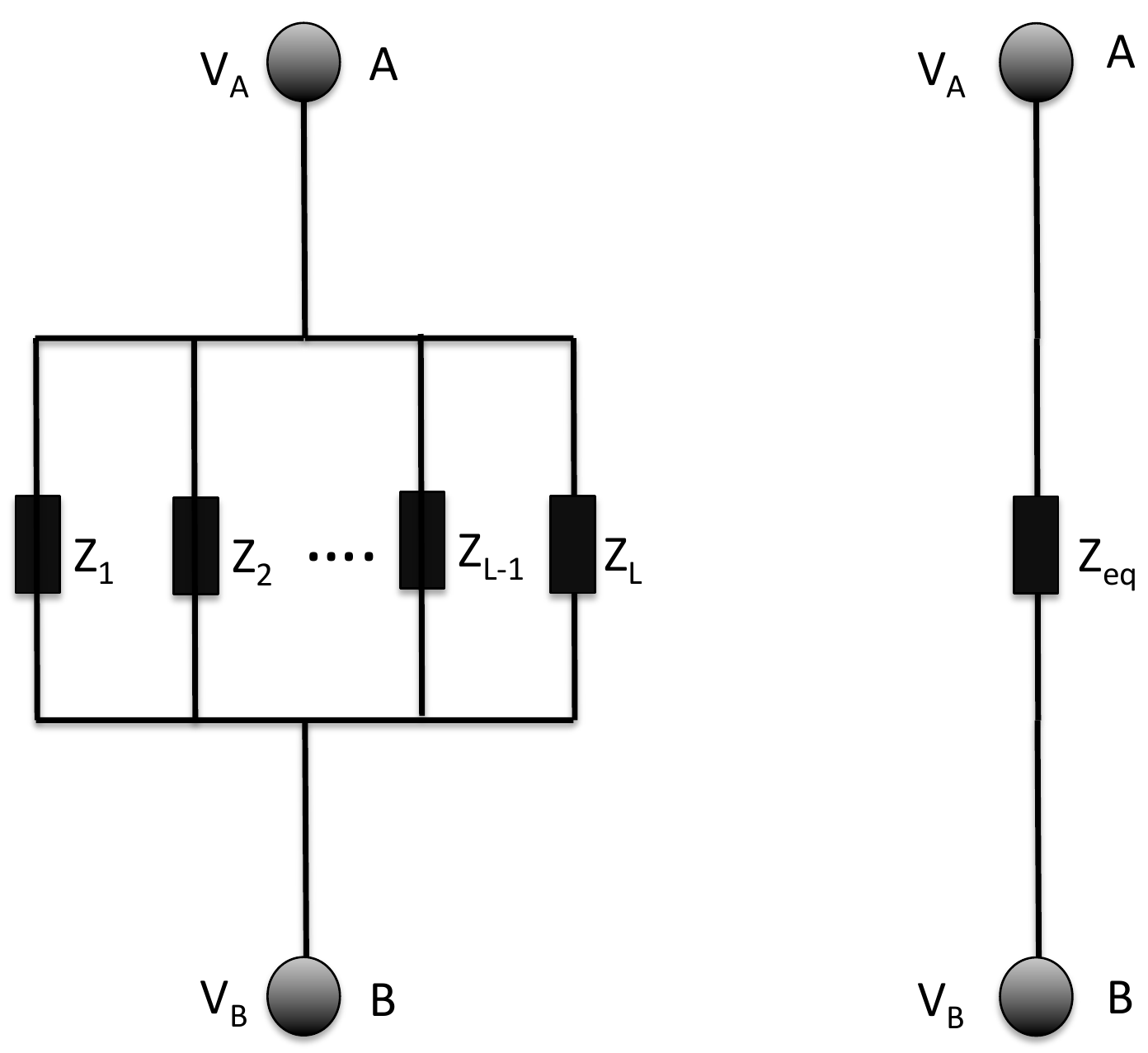}
\caption{The equivalent impedance transforms the multiple $L$ paths between nodes $A$ and $B$ with impedances $Z_{1},Z_{2},..,Z_{L}$ (on the left hand side) to a one single conceptual path with an impedance value $Z_{eq,AB}$ (on the right hand side).}
\label{fig:EqImpedance}
\end{figure}

Modelling power grid dynamics requires taking the power flow into account. In a dynamic power grid model, electric power flows through multiple paths. This precludes the existence of a distinct path (and subsequently the existence of the shortest path) between two nodes in a physical power grid. Yet, the concept of \emph{equivalent impedance} makes it possible to determine a distinct \emph{electrical path} between two nodes by conceptually replacing the multiple paths between two nodes with a single equivalent path. Fig.~\ref{fig:EqImpedance} illustrates the concept. 

\begin{figure*}
\begin{center}	
	\includegraphics[width=.35\textwidth]{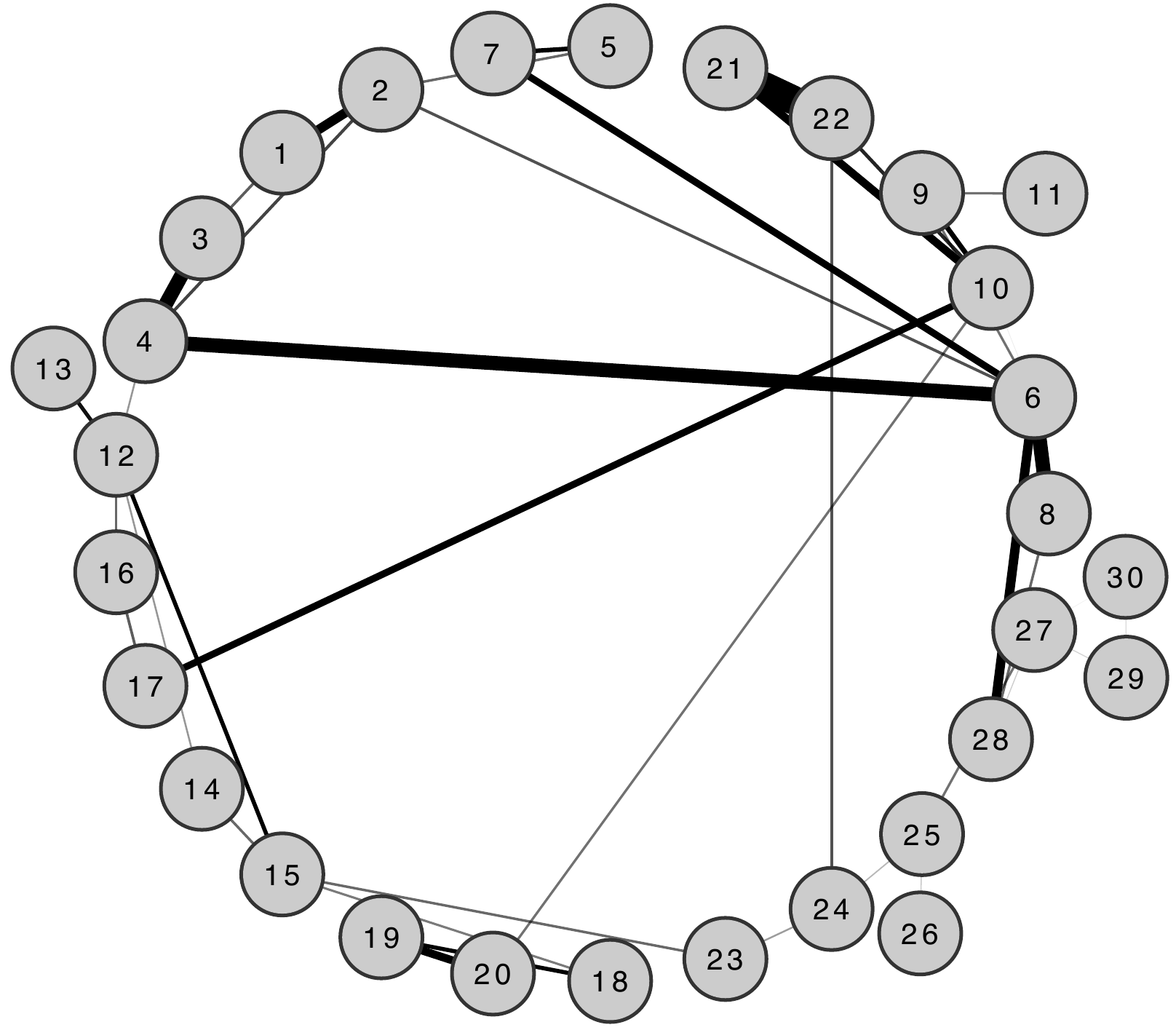}
	\label{fig:IEEE30Physical}
	\includegraphics[width=.31\textwidth]{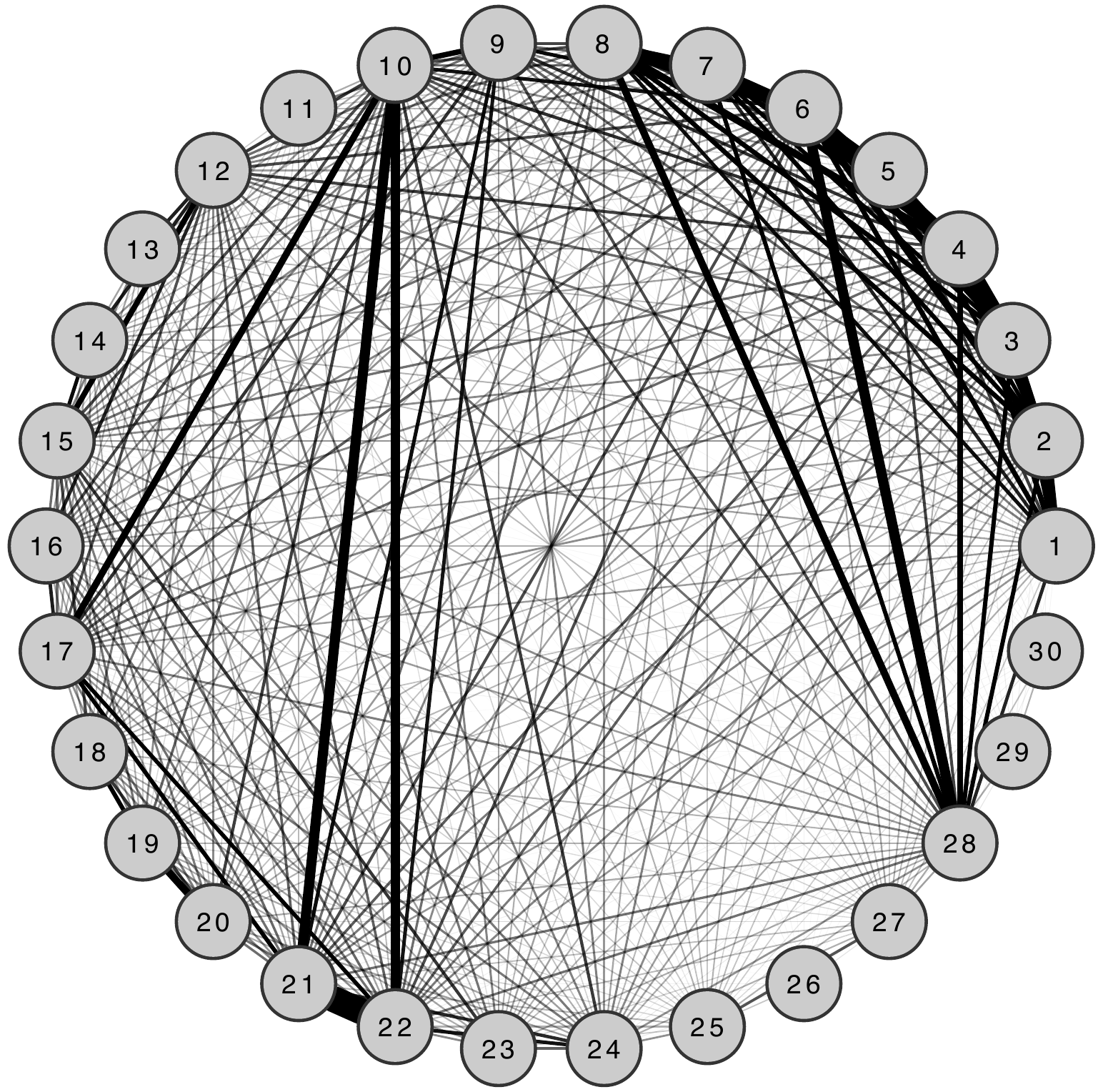}
	\label{fig:IEEE30Electrical}
\caption{The physical (left) and the electrical (right) topology of IEEE 30 power system. In the physical topology the conductances, and in the electrical topology the equivalent conductances (i.e.\xspace $1/Z_{eq}$), are used as weights for a better illustration. A relatively thicker and more visible line corresponds to a stronger connection/a shorter electric path length (i.e.\xspace a smaller effective resistance), while a relatively thinner and less visible line corresponds to a relatively weaker connection/ a larger electric path length (i.e.\xspace larger effective resistance).}
\label{fig:IEEE30}
\end{center}
\end{figure*}

The concept of electrical path length makes it possible to construct the \emph{electrical topology} of a power grid. An electrical topology of a power grid shows the electrical connections/path lengths (i.e.\xspace equivalent impedances) between buses, rather than the physical connections as a physical topology does. In a power grid, the actual "distance" between two nodes is the electrical path length, and not the physical distance (or number of lines). Accordingly, the electrical topology of a power grid governs the network dynamics rather than the physical topology. Fig.~\ref{fig:IEEE30} shows the physical and the electrical topology of the Institute of Electrical and Electronics Engineers (IEEE) 30 buses power system~\cite{TestCaseRef}. 

The initial concepts of "electrical distance" are discussed by power system researchers~\cite{Hines2008, Arianos2009, Bompard2009}, and also by mathematicians as \emph{effective resistance}~\cite{Klein1993, Ellens2011}. In a power grid, the effective resistance $R_{ij}$ between buses $i$ and $j$ equals the equivalent impedance between these buses. Hence, in the electrical topology in Fig.~\ref{fig:IEEE30}, the connections are the effective conductances ($1/R_{ij}$) between each pair of nodes. The \emph{effective graph resistance} $R_{G}$ of a power grid $G$ is an aggregate value of all effective resistances between any pair of nodes in a grid (see Eq.~\ref{EffGraphResistance}).      

The existence of parallel paths between two nodes in a physical power grid topology, and a homogeneous distribution of their impedance values result in a smaller effective resistance between these two nodes (i.e.\xspace a stronger connection between these nodes in the electrical topology). The number of parallel paths in the physical topology refers to the number of redundant (backup) paths. In case of a failure in one of the paths between two nodes, the power flow carried by the rendered path is distributed over the backup paths. Therefore, a higher number of backup paths implies a more robust network against cascading failures due to line overloads. On the other hand, a relatively more homogeneous distribution of the impedance values results in a relatively more homogeneous distribution of power flow over these parallel paths increasing the robustness of the power grid against cascading failures~\cite{Koc2013, Koc2013_2}. Therefore, a power grid with a relatively smaller effective graph resistance implies a relatively more robust power grid with respect to cascading failures. 

\section{\label{sec_Experimental Verification}Effective Graph Resistance and Power Grid Robustness: Experimental Verification}
This section verifies the potential of effective graph resistance to anticipate the cascading failures robustness of power networks. Experimental verification of the proposed robustness metric requires (i) creating synthetic test systems, (ii) determining the effective graph resistance of these synthetic test systems (i.e.\xspace theoretical results) and their robustness levels by simulations (i.e.\xspace experimental results that are considered as the ground truth), and (iii) quantifying the correlation between the theoretical results and experimental results to assess whether the effective graph resistance anticipates the power grid robustness with respect to cascading failures.

\subsection{\label{subsec_Test systems}Test systems}
The data required for the metric verification analysis includes the topology of a power grid (i.e.\xspace interconnection of buses with lines), reactance values of transmission lines, the types buses and their generation capacity and load values. Since the IEEE power systems provide all these data, they are considered as the reference systems and additional synthetic test systems are created based on these reference systems. 

The synthetic test systems are generated to have exactly the same properties as the reference IEEE test system (e.g.\xspace type and number of buses/lines with their interconnection and demand/generation capacity values) except for the effective graph resistance. In this way, in each test system, all the parameters that have an impact on the robustness (e.g.\xspace topology, generation/loading profile, and loading level) are fixed, except for the effective graph resistance. This assures that the difference in the robustness levels of these test systems is due to the change in the effective graph resistance of these grids, and makes possible to assess the impact of the effective graph resistance on the grid robustness.

When creating a new synthetic test system based on the IEEE reference system, it is crucial to have a different effective graph resistance and to keep other important aspects for the robustness unchanged. To achieve this, an arbitrary number of the links of the reference IEEE test systems is randomly chosen and the reactance values of these transmission lines are increased. For example, when creating a synthetic test system based on the IEEE 118 power system~\cite{TestCaseRef}, e.g.\xspace 4 transmission lines are randomly chosen: $l_{1}$, $l_{2}$, $l_{3}$, $l_{4}$. Then the reactance values of these lines (i.e.\xspace $x_{1}$, $x_{2}$, $x_{3}$, $x_{4}$) are doubled and a new synthetic test system is created with the new reactance values: $2x_{1}$, $2x_{2}$, $2x_{3}$, $2x_{4}$ for lines $l_{1}$, $l_{2}$, $l_{3}$, $l_{4}$ respectively. 

The new synthetic test system has an increased $R_{G}$ since the effective graph resistance is a monotonic increasing function of the individual reactance values in a network~\cite{Ellens2011}. It also has a different robustness level against cascading failures because the manipulated impedances affect the power flow distribution in the networks~\cite{Koc2013, Koc2013_2}. 

A second synthetic test system is created from the reference IEEE 118 power system by increasing the reactance values of the same lines $l_{1}$, $l_{2}$, $l_{3}$, $l_{4}$ by a factor of three resulting in the impedance values of $3x_{1}$, $3x_{2}$, $3x_{3}$, $3x_{4}$, respectively. In this way, the additional synthetic test systems are created.

 \subsection{\label{subsec_Robustness Levels}Robustness levels by effective graph resistance and by simulations}
The effective graph resistance supposedly quantifies the \emph{theoretical} robustness levels of the test systems against cascading failures. The $R_{G}$ of the test systems will be compared to the simulation-based robustness levels to assess whether $R_{G}$ truly anticipates the grid robustness. The computation of $R_{G}$ requires data about the admittance matrix of the systems (i.e.\xspace reactance values of the transmission lines and the topology of the grid). The effective graph resistance of each test system is calculated by Eq.~\ref{EffResistance} and Eq.~\ref{EffGraphResistance} in Sec.~\ref{sec_Graph Resistance Basics}. The \emph{experimental} robustness levels of created synthetic power grids are determined using simulations. MATCASC~\cite{Koc2013_3} (a MATLAB-based tool implementing cascading failure simulations in power grids) is used to simulate cascading failures and to quantify the grid robustness. Cascading failures by targeted attacks are simulated for different values of tolerance parameter $\alpha$ (i.e.\xspace for different loading levels). $DS$ metric is used to quantify the robustness of the test systems after the cascades subsides.
 
This paper uses an attack strategy based on the electrical node significance metric~\cite{Koc2013, Koc2013_2}. The electrical node significance is a contextual node centrality measure, specifically designed for power grids. The electrical node significance $\delta$ of a node \emph{i} is defined as the amount of power distributed by node \emph{i}, normalized by the total amount of the power that is distributed in the entire grid:

\begin{equation}\label{Delta}
\delta _{i}=\frac{P_{i}}{\sum_{j=1}^{N} P_{j}}
\end{equation}

\noindent where $P_{i}$ stands for total power distributed by node $i$ while $N$ refers to number of nodes in the network. This paper attacks a power grid by targeting \emph{the most heavily loaded outgoing link from the node with the highest electrical node significance} in the network. Removal of this link most likely results in the largest cascading failure in the power network~\cite{Verma2012}.

\begin{figure}
\centering
\includegraphics[scale=0.25]{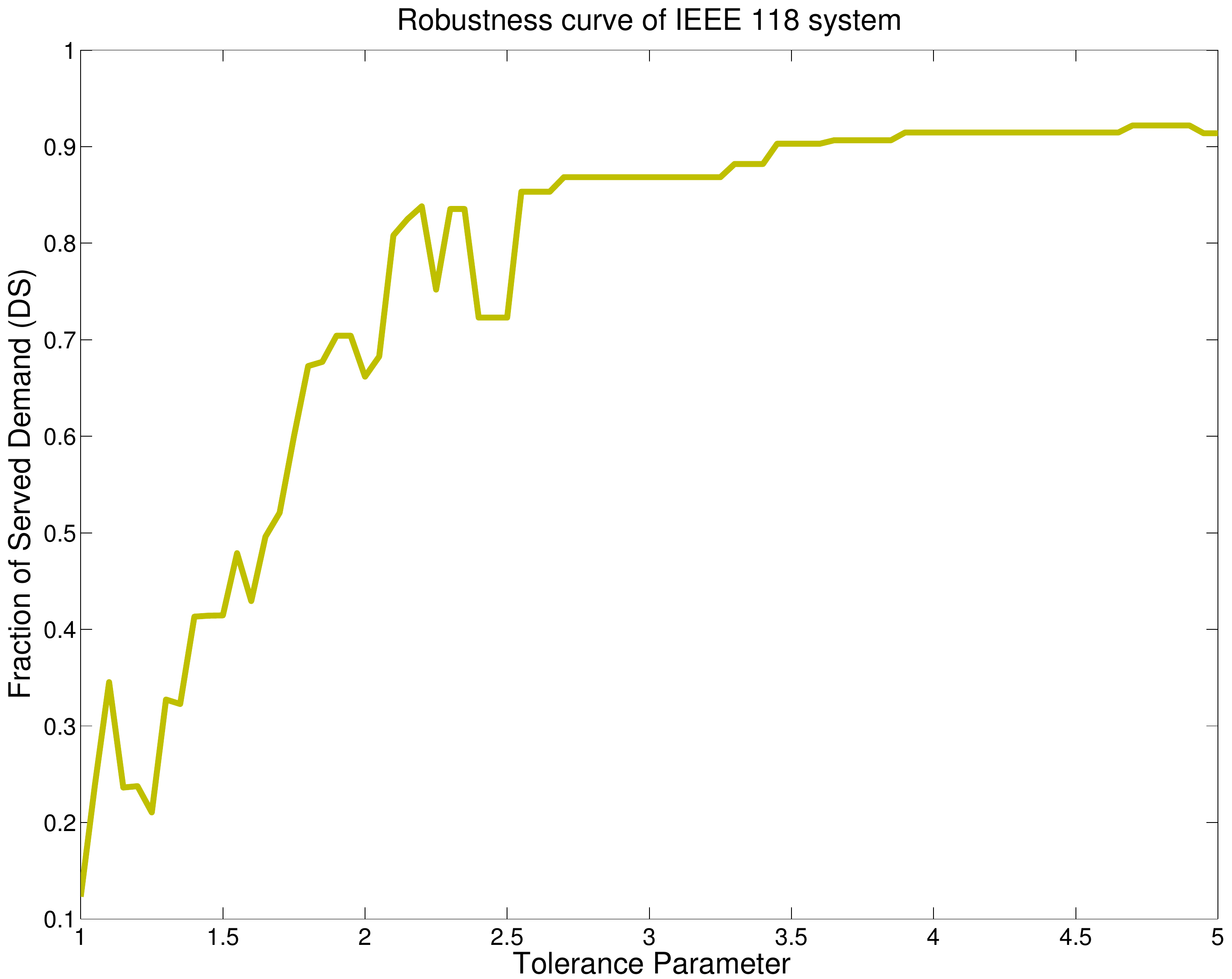}
\caption{The robustness curve for the IEEE 118 test system}
\label{fig:IEEE118RobCurve}
\end{figure}

 Assessing the robustness of a power grid against cascading failures by targeted attacks for an interval of tolerance parameters [$\alpha_{min}$, $\alpha_{max}$] subdivided by $\Delta_{\alpha}$ (i.e.\xspace a set of tolerance parameters [$\alpha_{min}$, $\alpha_{min}+\Delta_{\alpha}$, $\alpha_{min}+2\Delta_{\alpha}$, ..., $\alpha_{max}$]) results in the \emph{robustness curve} of the grid. For example, Fig.~\ref{fig:IEEE118RobCurve} shows the robustness curve for the IEEE 118 power system~\cite{TestCaseRef} obtained for [$\alpha_{min}$=$1$, $\alpha_{max}$=$5$] with $\Delta_{\alpha}$=$0.05$. The robustness curve in Fig.~\ref{fig:IEEE118RobCurve} suggests that, when the network is loaded by 20\% (i.e.\xspace $\alpha$=5), an attack results in the loss of almost 10\% of the total demand. However, an attack when $\alpha$=1.3 (i.e.\xspace $\emph{loading level} \simeq 77\%)$ results in collapse of the network and only 20\% of the total demand can be satisfied.

Traditionally, power grid researchers focus on the robustness of a grid for a specific tolerance level of the grid. However, this paper considers the robustness of a power grid over the whole spectrum of tolerance levels (i.e.\xspace loading levels), and defines mathematically the normalized area $r$ below the robustness curve as an empirical metric that quantifies the power grid robustness against cascading failures. The normalized area below a robustness curve is computed by a Riemann sum~\cite{Mieghem2006}:
\begin{equation}
\label{robOfRobCurve}
r=\frac{1}{(\alpha_{max}-\alpha_{min})}\sum_{\substack{i \in d}}{DS(\alpha_{i})\Delta_{\alpha}}
\end{equation}

%


\noindent where $DS(\alpha_{i})$ is the $DS$ after a cascading failure is triggered by an attack when the tolerance of the network is $\alpha_{i}$, and $d$ is the size of the set of tolerance parameters at which the robustness levels are determined. Because the maximum value of $DS$ is $1$, $(\alpha_{max}-\alpha_{min})$ refers to the maximum possible area below the robustness curve. Consequently, the factor $1/(\alpha_{max}-\alpha_{min})$ in Eq.~\ref{robOfRobCurve} normalizes the area below the robustness curve ensuring that $r$ has a value between 0 and 1. Evaluation of Eq.~\ref{robOfRobCurve} for the robustness curve of a grid results in the experimental robustness level of the grid with respect to cascading failures. On the other hand, the vulnerability of the grid is the complementary of the robustness of the grid and defined as:
\begin{equation}\label{vulnOfRobCurve}
v=1-r
\end{equation}

\noindent where $v$ is the measure of vulnerability of the system, also having a value between 0 and 1.

\subsection{\label{subsec_Assessment}Assessing the effectiveness of the effective graph resistance in anticipating power grid robustness}
\begin{figure*}
\centering
\includegraphics[scale=0.33] {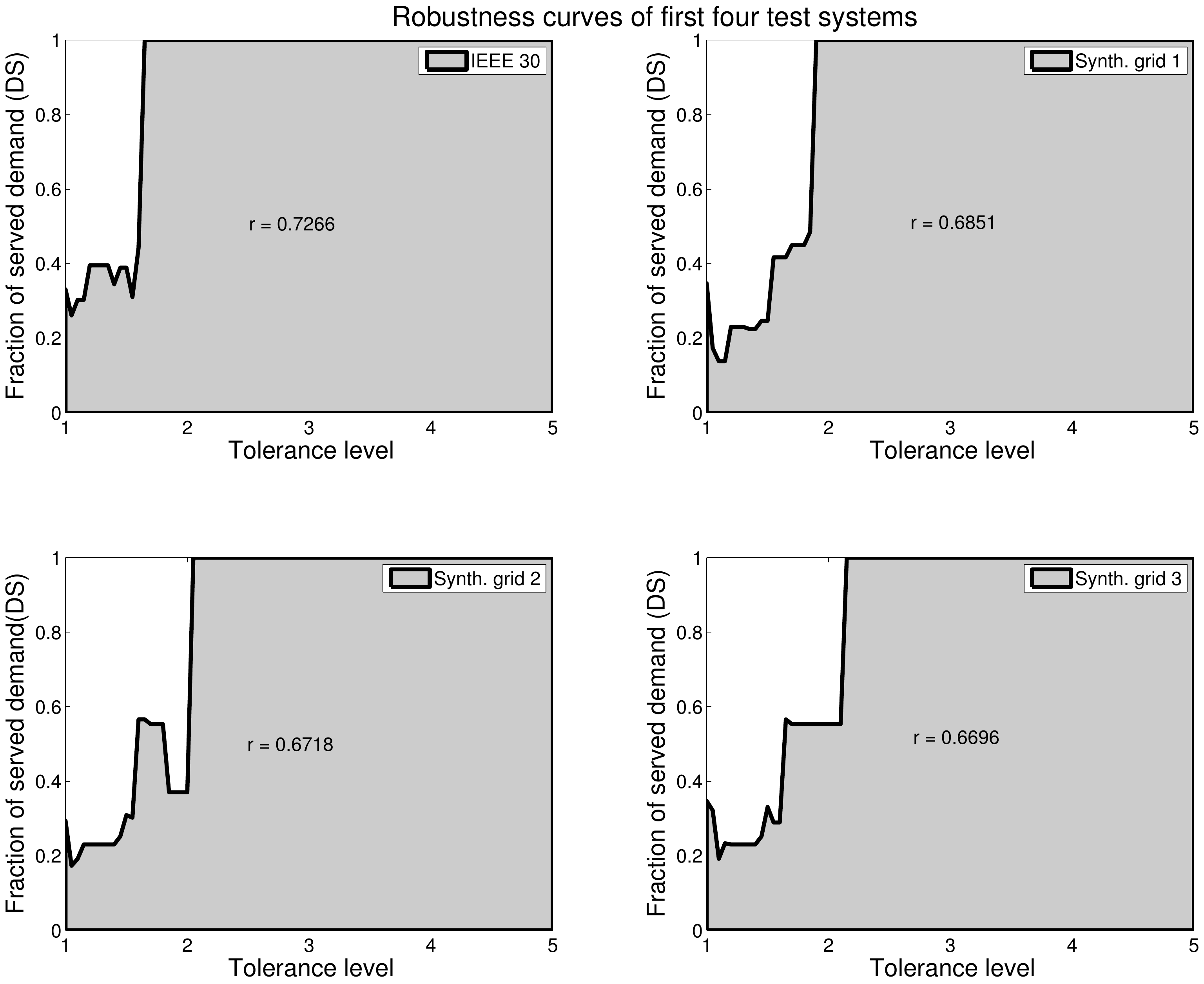}
\caption{The robustness curves for the IEEE 30 and first 3 synthetic test systems, and the corresponding robustness (i.e.\xspace $r$) values. The test systems have their effective graph resistances in an ascending order, while their robustness levels are step by step increasing.}
\label{fig:SynthGridsRobCurves}
\end{figure*}

To gain more insight into the impact of $R_{G}$ on the grid robustness, first a qualitative analysis is performed for a relatively small set of test systems. The IEEE 30 power system (see Fig.~\ref{fig:IEEE30}) is considered as a reference system. 4 lines are chosen randomly: $l_{2}$ (connecting nodes $1$ and $3$), $l_{14}$ (connecting nodes $9$ and $10$), $l_{38}$ (connecting nodes $27$ and $30$), $l_{41}$ (connecting nodes $6$ and $28$), and based on the methodology explained in Sec.~\ref{subsec_Test systems}, three additional test systems are created. The robustness of these systems are determined by $R_{G}$ and by $r$. Table~\ref{tab:5 networks} provides the $R_{G}$ and $r$ values while Fig.~\ref{fig:SynthGridsRobCurves} shows the robustness curves of these test systems. 

\begin{table}
\centering
\caption{Effective graph resistance ($R_{G}$) and simulation-based robustness levels ($r$) of IEEE 30 and 3 synthetic test systems}
\label{tab:5 networks}
\begin{tabular}{l c c } \hline\hline
 						 & $R_{G}$	& $r (\%)$\\
\hline
IEEE 30				  & 151.86	& 72.66\\
Synthetic grid 1 & 162.71	& 68.51\\
Synthetic grid 2 & 169.43	& 67.18\\
Synthetic grid 3 & 174.07	& 66.96\\
\hline \hline
\end{tabular}
\end{table}

The effective graph resistance is a monotonic increasing function of the individual impedance (consisting of resistance and reactance) values in the grid~\cite{Ellens2011}. Accordingly, in Table~\ref{tab:5 networks}, $R_{G}$ increases as a result of the increase of individual reactance values. This indicates that the typical average electrical path length between the buses in the synthetic grids is increasing incrementally making these grids relatively loosely coupled. On the other hand, as a result of the increase of reactance values in these synthetic grids, $r$ becomes step by step smaller: synthetic grid I is more robust compared to synthetic grid II, synthetic grid II is more robust compared to synthetic grid III, and so forth. Hence, the theoretical results (i.e.\xspace $R_{G}$) are evidently in line with the simulation results, and suggest that increasing the effective graph resistance of a power grid makes it less robust (i.e.\xspace more vulnerable) against cascading failures by targeted attacks. Fig.~\ref{fig:SynthGridsRobCurves} visualizes the impact of $R_{G}$ on grid robustness: an increase in $R_{G}$ makes the power grid robustness smaller.

\begin{figure*}
\begin{center}	
	\includegraphics[width=.46\textwidth]{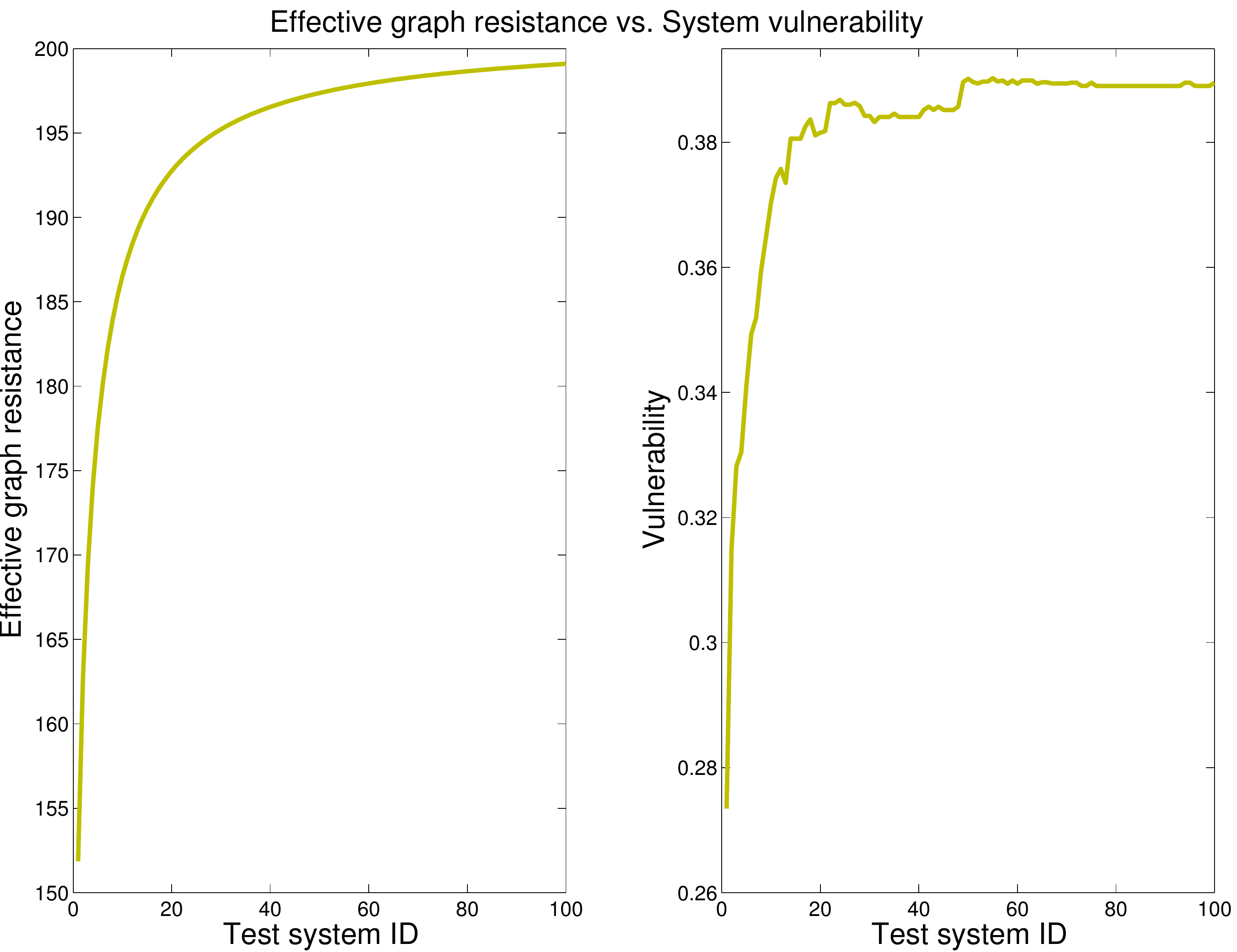}
	\label{fig:IEEE30TheoretVSExperRobLevels_ByR}
	\includegraphics[width=.45\textwidth]{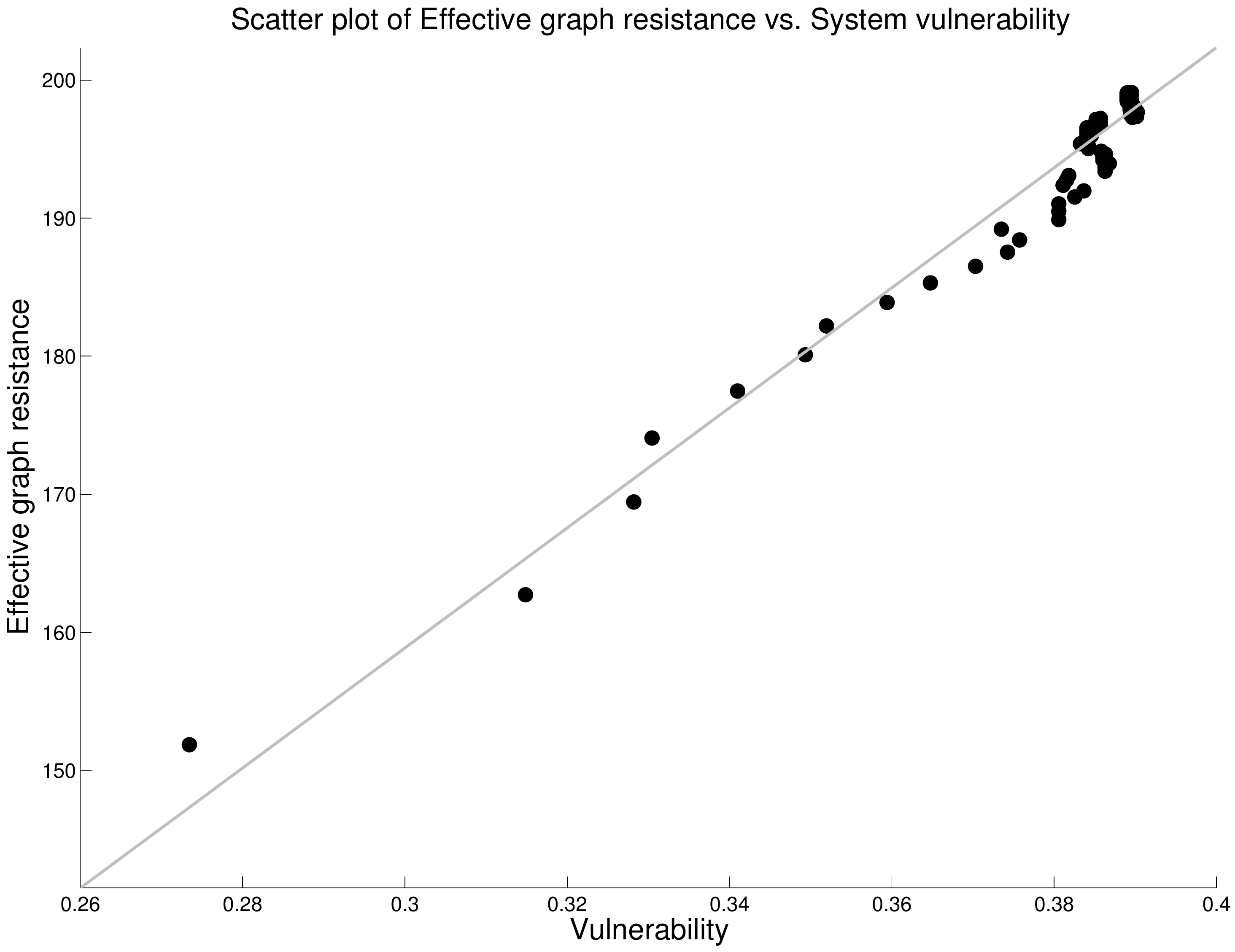}
	\label{fig:Scatter}
\caption{Experimental verification of the effective graph resistance for IEEE 30 test system. For 100 synthetic test systems, on the left hand side the effective graph resistance and the vulnerability values are plotted. On the right hand side a scatter plot visualizes the correlation between effective graph resistance and system vulnerability. Results show that the effective graph resistance captures the system vulnerability accurately.}
\label{fig:IEEE30Verification}
\end{center}
\end{figure*}

A quantitative assessment of the impact of the effective graph resistance on the power grid robustness requires the robustness analysis for a larger set of networks. Accordingly, the small set of 4 test systems (see Table~\ref{tab:5 networks}) is expanded to a set of 100 test systems. The effective graph resistances and the empirical robustness levels (i.e.\xspace $r$) of these networks are determined. The correlation level between these theoretical and experimental robustness levels is quantified: the linear correlation coefficient between $R_{G}$ and $r$ is over -90\%. This nearly perfect anti-correlation level suggests that the effective graph resistance anticipates the power grid robustness with respect to cascading failures accurately. Fig.~\ref{fig:IEEE30Verification} plots $R_{G}$, $v$, and scatter diagram of $R_{G}$ and $v$. Results in Fig.~\ref{fig:IEEE30Verification} visualizes how the effective graph resistance captures the power grid robustness. The vulnerability is plotted rather than the experimental robustness $r$ for better illustration of coherence between the theoretical and experimental results.


Alternating the reactance values in a grid causes a different power flow distribution in the grid and, in turn, results in a different level of robustness against cascading failures~\cite{Koc2013, Koc2013_2}. Fig.~\ref{fig:IEEE30Verification} shows that the vulnerability $v$ reflects the variations in $R_{G}$ of the power grids accurately: a fast increase in the grid effective graph resistance results in a fast increase in the grid vulnerability (i.e.\xspace fast drop in grid robustness ), while a slight increase in the grid effective graph resistance causes a slight increase in the grid vulnerability (i.e.\xspace a slight drop in the grid robustness). 

\begin{figure}
\centering
\includegraphics[scale=0.28]{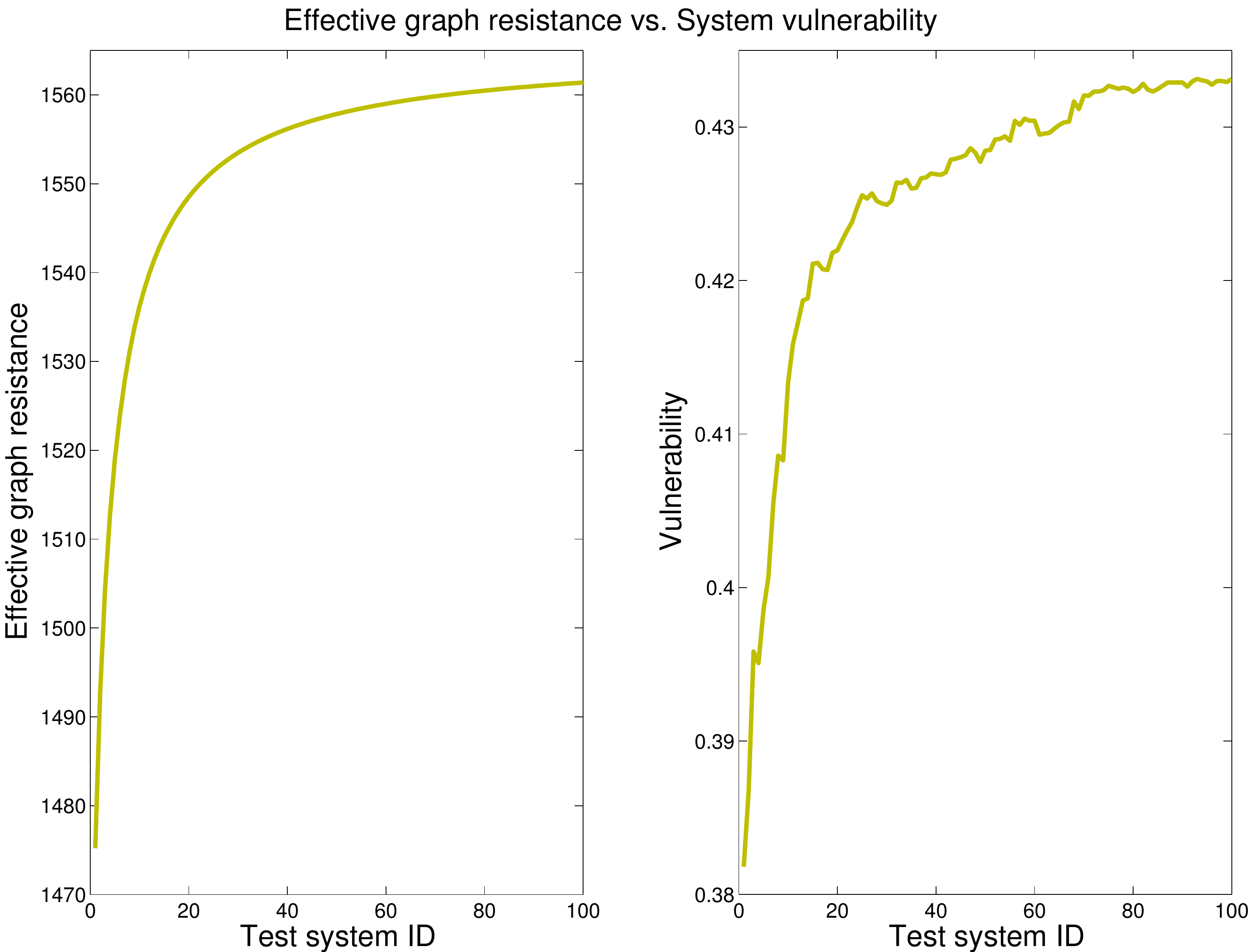}
\caption{Experimental verification of the effective graph resistance for IEEE 118 test system. For 100 synthetic test systems, the effective graph resistance (on the left hand side), and the vulnerability values (on the right hand side) are plotted.}
\label{fig:seven}
\end{figure}

To investigate the validity of the impact of $R_{G}$ on power grid robustness, the robustness analysis is performed for the larger IEEE 118 power systems. Again four lines are randomly chosen: $l_{31}$ (connecting buses 23 to 25), $l_{35}$ (connecting buses 28 to 29), $l_{93}$ (connecting buses 59 to 63), and $l_{175}$ (connecting buses 109 to 110). Furthermore, 99 additional test systems are created. For these test systems, $R_{G}$ is computed and the experimental robustness levels are obtained by $r$ (and grid vulnerability by $v$). Fig.~\ref{fig:seven} shows the $R_{G}$ and $v$ values for the 100 test systems. The correlation level between $R_{G}$ and $r$ resides again over -90\%.

The experimental results (see Fig.~\ref{fig:IEEE30Verification} and Fig.~\ref{fig:seven}) show that effective graph resistance captures power grid robustness very accurately and suggests that an increased effective graph resistance results in a less robust power grid against cascading failures, as proposed in Sec.~\ref{sec_Graph Resistance}.

\section{\label{sec_Use cases}Use Case: Application for Power Grids}
The aptitude of effective graph resistance to relate a power grid topology to its robustness can be exploited in different ways, including designing robust networks from scratch and identifying critical components in a power grid. Furthermore $R_{G}$ also acts as a measure based on which a grid topology is optimized to maximize the grid robustness. The next section focusses on optimization of the IEEE 118 power system topology for a higher level of robustness against cascading failures. 

 \subsection{ \label{subsec_Upgrading an existing power grid}Upgrading IEEE 118 power system to improve grid robustness}
As a response to blackouts, additional transmission lines are placed to increase the power grid robustness. Determining the right pair of buses to connect in order to maximize the grid robustness is the challenge\footnote{This section focuses on maximizing the grid robustness rather than minimizing the economic cost of placing a transmission line.}. A solution to this problem requires (i) determining all possible additional transmission lines (i.e.\xspace candidate transmission lines), and (ii) assessing the impact of each candidate transmission line on the effective graph resistance.  

For a grid consisting of N buses and L transmission lines, an additional line can be placed between any unconnected pair of nodes. Therefore, there are in total $\frac{N(N-1)}{2}-L$ number of lines that can be added to an existing power grid. 

To assess the impact of an additional line on the effective graph resistance (i.e.\xspace on the grid robustness), this section deploys an analogous approach to the one given in~\cite{Latora2005, Mieghem2011_2}: the optimum location for an additional line is determined by assessing its relative change on the effective graph resistance value of the original power grid. Because the effective graph resistance is a monotonic decreasing function of the number of lines~\cite{Ellens2011} in a grid, adding a transmission line to an existing grid decreases the effective graph resistance of the grid. Therefore, the optimum location of an additional line is determined based on the relative decrease in $R_{G}$ that is caused by adding the line $l$:
\begin{equation}\label{CriticalityUpgrade}
\Delta R_{G}^{l}=\frac {R_{G}-R_{G+l}}{R_{G}} 
\end{equation}

\noindent where $G$, $R_{G}(G+l)$ is the effective graph resistance of the improved grid, and $\Delta R_{G}^{l}$ is the relative decrease in the effective graph resistance of $G$ as a result of adding line $l$.   

\begin{figure}
\centering
\includegraphics[scale=0.30]{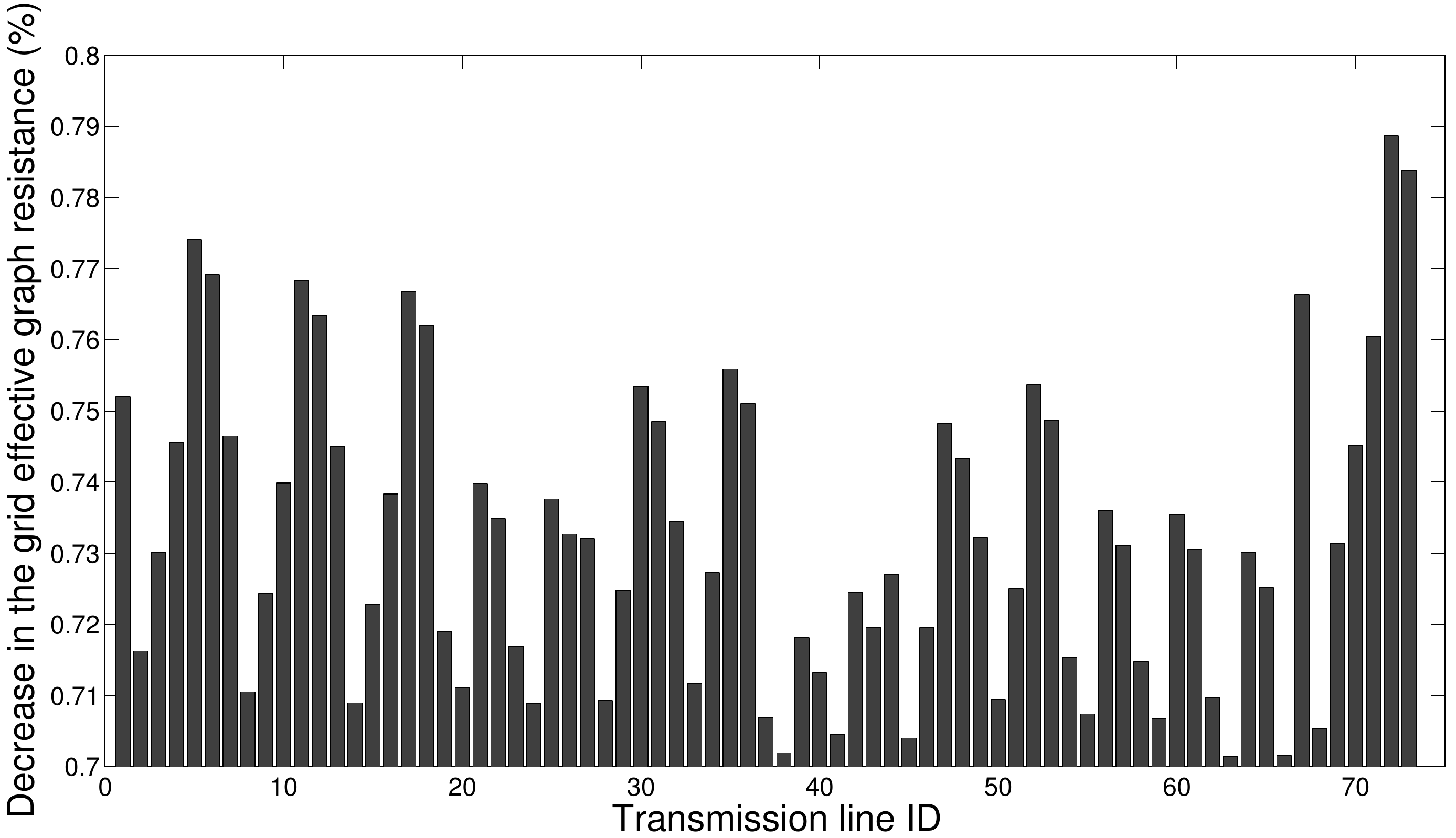}
\caption{Relative decrease in the effective graph resistance as a result of adding a line for IEEE 118 test case. From 6724 total lines, 74 lines whose addition to IEEE 118 buses power system results in a $R_{G}$ decrease of 0.7\% or higher.}
\label{fig:IEEE118LineAdditionAnalysis}
\end{figure}

The IEEE 118 power system is considered as a test case and the optimum location for an additional line to upgrade the power system is investigated. The IEEE 118 test system has 118 buses and 186 transmission lines. Hence, there are 6724 possible lines to be added. The reactance value of a candidate transmission line is assumed to be the average of all transmission lines in the IEEE 118 test system. For each of these candidate lines, Eq.~\ref{CriticalityUpgrade} is evaluated and its impact on the effective graph resistance is determined. From these 6724 lines, Table~\ref{tab:10 influential lines} shows the top 10 most influential transmission lines to be added to IEEE 118 power system, while Fig.~\ref{fig:IEEE118LineAdditionAnalysis} plots the lines that have an impact higher than 0.7\%. 

The results in Table~\ref{tab:10 influential lines} and Fig.~\ref{fig:IEEE118LineAdditionAnalysis} can be used as starting point for an economic analysis. Since the impact of adding a line is not very different for the top ten candidates (see Table~\ref{tab:10 influential lines} and Fig.~\ref{fig:IEEE118LineAdditionAnalysis}) a cost function can be used to determine which of these lines provides a significant increase in grid robustness at an acceptable or minimal cost.

\begin{table}
\centering
\caption{The top 10 most influential lines to add in IEEE 118 system}
\label{tab:10 influential lines}
\begin{tabular}{l c} \hline
 	Line ID 			  				& $\Delta R_{G}^{l}$(\%)\\
\hline \hline
$l_{111-117}$			  		& 0.788\\
$l_{112-117}$			  		& 0.783\\
$l_{1-111}$ 						& 0.774\\
$l_{1-112}$ 						&0.769\\
$l_{2-111}$ 						& 0.768\\
$l_{3-111}$			  			& 0.767\\
$l_{87-117}$			  		& 0.765\\
$l_{2-112}$ 						& 0.763\\
$l_{3-112}$ 						&0.761\\
$l_{110-117}$ 				& 0.760\\
\hline\hline
\end{tabular}
\end{table}

Table~\ref{tab:10 influential lines} shows that the largest decrease in the effective graph resistance is achieved by adding a transmission line between nodes 111 and 117 (line ID 73 in Fig.~\ref{fig:IEEE118LineAdditionAnalysis}). Placing a line between these nodes results in a decrease of nearly 0.79\%  in the effective graph resistance of the original power system.

\begin{figure}
\centering
\includegraphics[scale=0.35]{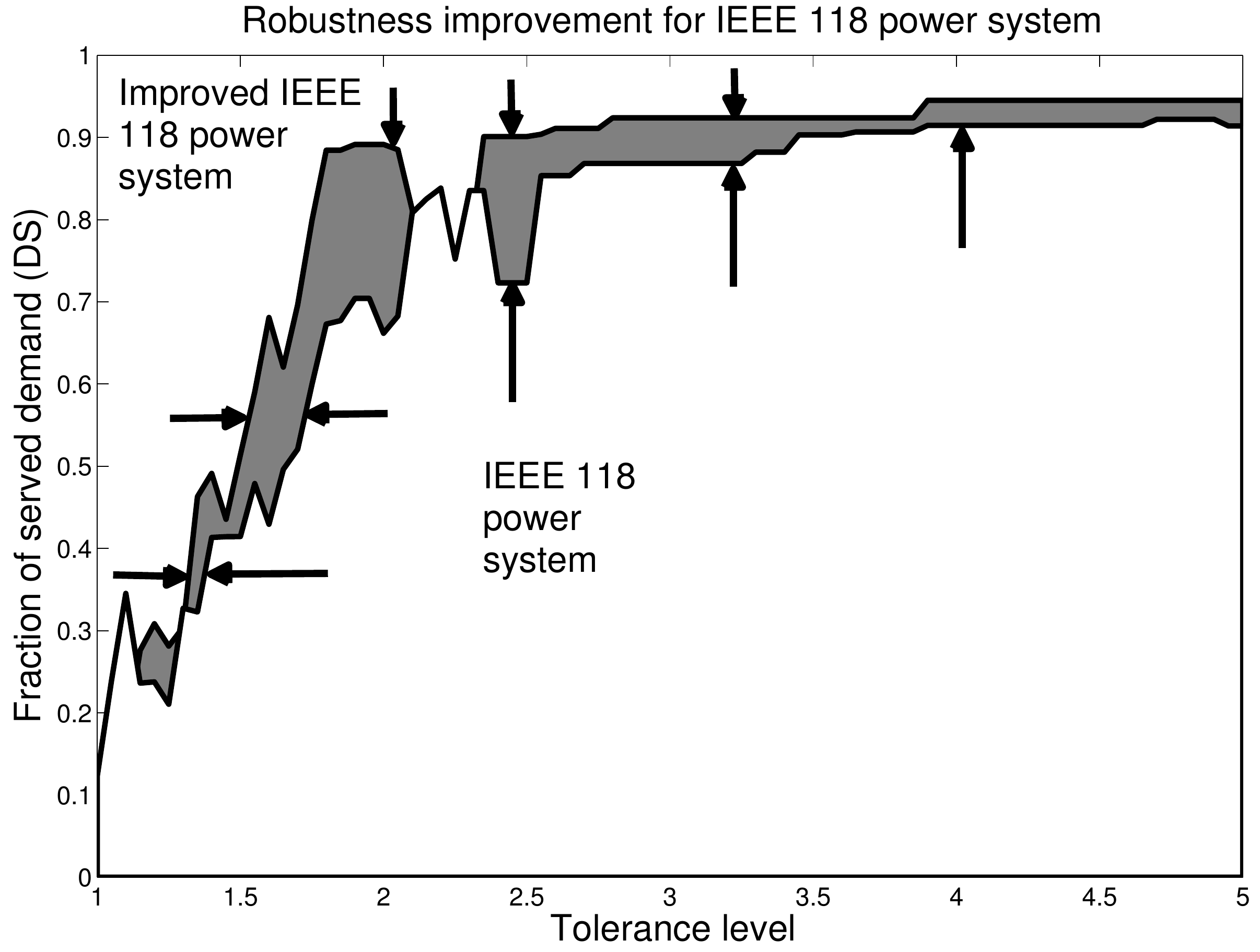}
\caption{Robustness curves of IEEE 118 and improved IEEE 118 power systems. The difference (i.e.\xspace robustness improvement) is highlighted.}
\label{fig:RobCurvesOfIEEE118AndImprovedIEEE118}
\end{figure}

To assess the impact of placement of the additional line $l_{111-117}$ on the grid robustness, the robustness curve of the improved topology, and its robustness $r$ are determined. The improved grid has a robustness $r$=0.6570 which is an increase by 6.7\% compared to the original grid robustness (i.e.\xspace 0.6182). Fig.~\ref{fig:RobCurvesOfIEEE118AndImprovedIEEE118} plots the robustness curves of both topologies, and highlights the improvement in the grid robustness (shaded area in grey in Fig.~\ref{fig:RobCurvesOfIEEE118AndImprovedIEEE118}).  

\section{\label{sec_Conclusion}Discussion and Conclusion}
This paper proposes the \emph{effective graph resistance} $R_{G}$ as a metric to assess the impact of the power grid topology $G$ on the robustness against cascading failures due to line overloads by targeted attacks. The effective graph resistance quantifies a power grid robustness by taking the number of backup paths and their ability to accommodate power flows into account. The experimental verification shows that $R_{G}$ anticipates the grid robustness against cascading line failures accurately. Experimental results suggest that increasing the effective graph resistance of a power grid results in a decreased grid robustness against cascading failures by targeted attacks. The proposed metric is used to optimize the topology of the IEEE 118 power system to improve its robustness. Results show that adding a single line to IEEE 118 power system based on the effective graph resistance analysis improves the grid robustness by 6.7\%.

\vspace*{\fill}

\end{document}